\documentstyle[12pt]{article}
\newlength{\extraspace}
\newlength{\extraspaces}
\newcommand{\be}{\begin{equation}
\addtolength{\abovedisplayskip}{\extraspaces}
\addtolength{\belowdisplayskip}{\extraspaces}
\addtolength{\abovedisplayshortskip}{\extraspace}
\addtolength{\belowdisplayshortskip}{\extraspace}}
\newcommand{\ee}{\end{equation}}

\newcommand{\ba}{\begin{eqnarray}
\addtolength{\abovedisplayskip}{\extraspaces}
\addtolength{\belowdisplayskip}{\extraspaces}
\addtolength{\abovedisplayshortskip}{\extraspace}
\addtolength{\belowdisplayshortskip}{\extraspace}}
\newcommand{\ea}{\end{eqnarray}}

\newcommand{\nonu}{\nonumber \\[.5mm]}
\newcommand{\A}{&\!\!\!}

\newcommand{\newsection}[1]{
\vspace{7mm} \pagebreak[3] \addtocounter{section}{1}
\setcounter{subsection}{0} \setcounter{footnote}{0}
\begin{center}
{\large {\bf \thesection. #1}}
\end{center}
\nopagebreak
\medskip
\nopagebreak \hspace{3mm}}

\begin{document}

\begin{center}
{{\bf Cylindrically Symmetric Solution  in Teleparallel Theory of
Gravitation}}
\end{center}
\centerline{ Gamal G.L. Nashed}

\bigskip

\centerline{{\it Mathematics Department, Faculty of Science, Ain
Shams University, Cairo, Egypt }}

\bigskip
 \centerline{ e-mail:nasshed@asunet.shams.eun.eg}

\hspace{2cm}
\\
\\
\\
\\
\\
\\
\\
\\

The field equations of a special class of teleparallel  theory of
gravitation and electromagnetic fields have been applied to tetrad
space having cylindrical symmetry with four unknown functions of
radial coordinate $r$ and azimuth angle $\theta$. The vacuum
stress-energy momentum tensor with one assumption concerning  its
specific form generates one non-trivial  exact analytic solution.
This  solution is characterized by a constant magnetic field
parameter $B_0$. If $B_0=0$ then, the solution will reduces to the
flat spacetime. The energy content is calculated using the
superpotential given in the framework of teleparallel geometry.
The energy contained in a sphere is found to be different from the
pervious results.

\newpage
\begin{center}
\newsection{\bf Introduction}
\end{center}

Theories of gravity based on the geometry of distance parallelism
\cite{YJ}$\sim$\cite{Jm} are commonly considered as the closest
alternative to the general relativity theory. Teleparallel gravity
models posses a number of attractive features both from the
geometrical and physical viewpoints. Teleparallelism is naturally
formulated by gauging external (spacetime) translation and
underlain the Weitzenb$\ddot{o}$ck spacetime characterized by the
metricity condition and by the vanishing of the curvature tensor.
Translations are closely related to the group of general
coordinate transformations which underlies general relativity.
Therefore, the energy-momentum tensor represents the matter source
in the field equations of tetradic theories of gravity like in
general relativity.

An important point of teleparallel gravity is that it corresponds
to a gauge theory for the translation group. As a consequence of
translations, any gauge theory including these transformations
will differ from the usual internal gauge models in many ways, the
most significance being the presence of the tetrad field. The
tetrad field can be used to define a linear Weitzenb$\ddot{o}$ck
connection, from which torsion can be defined but no curvature.
Also tetrad field can be used to define a Riemannian metric,  in
terms of which Live-Civita connection is constructed. It is
important to keep in mind that torsion and curvature are
properties of a connection and many different connection can be
define on the same manifold.

Teleparallel theories of gravity have been considered long time
ago in connection with attempts to define the energy of
gravitational field \cite{JJTK, M4}. It is clear from the
properties of the solutions of Einstein field equation of an
isolated system that a consistent expression for the energy
density  of the gravitational field would be given in terms of
second order derivatives of the metric tensor. It is well known
that there exists no covariant, nontrivial expression constructed
out of the metric tensor, both in three and four dimensions that
contain such derivatives. However, covariant expressions that
contain second order derivatives of the tetrad fields are
feasible. Thus it is legitimate to conjecture that the
difficulties regarding the problem of defining the gravitational
energy-momentum is related to the geometrical description of the
gravitational field rather than being an intrinsic drawback of the
theory \cite{SL}.

The problem of defining a consistence and unequivocal expression
for the energy of the gravitational and electromagnetic fields is
still open and an important question in general relativity. It is
well know that the principle of equivalence has led to the belief
that the gravitational energy cannot be localized \cite{RC,MT}.
However, the argument based on this principle regarding the
nonlocalizibility of gravitational  energy is controversial and
not generally accepted \cite{SL}. Therefore, it is legitimate to
conjecture that the difficulties associated to the problem of
defining the gravitational energy is related to the geometrical
description of the gravitational field, rather being an intrinsic
nuisance of the theory \cite{SL}. The first attempts to define the
energy of the gravitational field were based on pseudotensors
\cite{LL}, which make use of coordinate dependent expressions.
More recently the idea of quasi-local energy, i.e., energy
associated to a closed spacelike two surface, in the context of
the Hilbert-Einstein  action integral, has emerged as a tentative
description of the gravitational energy \cite{BY}.

Teleparallel  Theories of gravity, whose basic entities are the
tetrad field $e_{a \mu}$ (a and $\mu$ are SO(3,1) and space-time
indices, respectively) have been considered long time ago by M\o
ller \cite{M4} in connection also with attempts to define the
energy of the gravitational field. Teleparallel theories of
gravity are defined on the  Weitzenb$\ddot{o}$ck spacetime
\cite{Wr}, which is endowed with the affine connection. The
curvature tensor constructed out of this connection vanishes
identically. This connection defines a space-time with an absolute
parallelism or teleparallelism of vector fields \cite{HS1}. In
this geometrical framework the gravitational effects  are due to
the torsion tensor corresponding to the above mentioned
connection.

As stated above that the calculations of energy within the
framework of general relativity theory have some problems
\cite{M4}. It is the aim of the present work to study a tetrad
having cylindrical  symmetry and apply it to the field equation of
gravitation and electromagnetic. Solving the resulting non linear
differential equation an exact solution is obtained. We then,
calculate the energy using the superpotential of Mikhail et
 al. \cite{MWHL}.

In section 2, we give a brief review of the gravitational and
electromagnetic theory. The tetrad having cylindrical symmetry is
applied to the field equations of the gravitational and
electromagnetic theory in section 3. The solution of the resulting
field equations is also given in section 3. The singularities of
this solution are given in section 4. In section 5, the
calculations of energy using the energy-momentum complex derived
by Mikhail et al. are given.  Section 6 is devoted for the
discussion and a summary of the results. A comparison between the
energy calculated and that of general relativity is also given in
section 6.

\newsection{The tetrad theory of gravitation and
electromagnetism}

In teleparallel theory of gravity and electromagnetism the
spacetime is represented by the  Weitzenb$\ddot{o}$ck  manifold
$W^4$ of distance parallelism. This theory naturally arises within
the gauge approach based on the group of the spacetime
translations. Accordingly, at each point of this manifold, a gauge
transformation is defined as a local translation of the tangent-
space coordinate \cite{Vt}, \[ x^a\rightarrow x'^a=x^a+b^a,\]
where $b^a=b^a(x^\mu)$ are the transformation parameters. for an
infinitesimal transformation we have \[ \delta x^a=\delta b^c P_c
x^a\]  with $\delta b^a$ the infinitesimal parameters, and
$P_a=\partial_a$ the generators of translations. Denoting the
translational gauge potential by ${\Lambda^a}_\mu,$ the gauge
covariant derivative for a scalar field $\Phi(x^\mu)$ reads
\cite{AP} \be D_\mu \Phi= {e^a}_\mu \partial_a \Phi,\ee where \be
{e^a}_\mu=\partial_\mu x^a+{\Lambda^a}_\mu,\ee is the tetrad field
which satisfies the orthogonality condition \be {e^a}_\mu
{e^\nu}_a={\delta^\nu}_\mu.\ee This nontrivial tetrad field
 induces a teleparallel structure on spacetime which is directly
 related to the presence of the gravitational field, and the
 Riemannian metric arises as  \be
g^{\mu \nu} \stackrel{\rm def.}{=} {e_i}^\mu {e_i}^\nu. \ee The
gravitational Lagrangian ${\it L}$ is an invariant constructed
from $\gamma_{\mu \nu \rho}$ and $g^{\mu \nu}$, where $\gamma_{\mu
\nu \rho}$ is the contorsion tensor given by \be \gamma_{\mu \nu
\rho} \stackrel{\rm def.}{=} e_{i \ \mu }e_{i \nu; \ \rho} , \ee
where the semicolon denotes covariant differentiation with respect
to Christoffel symbols. The most general Lagrangian density
invariant under parity operation is given by the form \cite{Mo1}
\be {\cal L} \stackrel{\rm def.}{=} (-g)^{1/2} \left( \alpha_1
\Phi^\mu \Phi_\mu+ \alpha_2 \gamma^{\mu \nu \rho} \gamma_{\mu \nu
\rho}+ \alpha_3 \gamma^{\mu \nu \rho} \gamma_{\rho \nu \mu}
\right), \ee where \be g \stackrel{\rm def.}{=} {\rm det}(g_{\mu
\nu}),
 \ee
 and
$\Phi_\mu$ is the basic vector field defined by \be \Phi_\mu
\stackrel{\rm def.}{=} {\gamma^\rho}_{\mu \rho}. \ee Here
$\alpha_1, \alpha_2,$ and $\alpha_3$ are constants determined by
M\o ller
 such that the theory coincides with general relativity in the weak fields \cite{Mo1}:

\be \alpha_1=-{1 \over \kappa}, \qquad \alpha_2={\lambda \over
\kappa}, \qquad \alpha_3={1 \over \kappa}(1-\lambda), \ee where
$\kappa$ is the Einstein constant and  $\lambda$ is a free
dimensionless parameter\footnote{Throughout this paper we use the
relativistic units, $c=G=1$ and
 $\kappa=8\pi$.}. The same
choice of the parameters was also obtained by Hayashi and Nakano
\cite{HN}.

The electromagnetic Lagrangian  density ${\it L_{e.m.}}$ is given
by  \cite{KT} \be {\it L_{e.m.}}=-\displaystyle{1 \over 4} g^{\mu
\rho} g^{\nu \sigma} F_{\mu \nu} F_{\rho \sigma}, \ee where
$F_{\rho \sigma}$ is given by\footnote{Heaviside Lorentz
rationalized unites will be used throughout this paper} \be F_{\mu
\nu}=
\partial_\mu A_\nu-\partial_\nu A_\mu, \ee where $A_\mu$ is the
electromagnetic potential.

The gravitational and electromagnetic field equations for the
system described by ${\it L_G}+{\it L_{e.m.}}$ are the following

 \be G_{\mu \nu} +H_{\mu \nu} =
-{\kappa} T_{\mu \nu}, \ee \be K_{\mu \nu}=0, \ee \be
\partial_\nu \left( \sqrt{-g} F^{\mu \nu} \right)=0 \ee
where the Einstein tensor $G_{\mu \nu}$ is defined by \be G_{\mu
\nu}=R_{\mu \nu}-{1 \over 2} g_{\mu \nu} R. \ee Here $H_{\mu \nu}$
and $K_{\mu \nu}$ are given by \be H_{\mu \nu} \stackrel{\rm
def.}{=} \lambda \left[ \gamma_{\rho \sigma \mu} {\gamma^{\rho
\sigma}}_\nu+\gamma_{\rho \sigma \mu} {\gamma_\nu}^{\rho
\sigma}+\gamma_{\rho \sigma \nu} {\gamma_\mu}^{\rho \sigma}+g_{\mu
\nu} \left( \gamma_{\rho \sigma \lambda} \gamma^{\lambda \sigma
\rho}-{1 \over 2} \gamma_{\rho \sigma \lambda} \gamma^{\rho \sigma
\lambda} \right) \right],
 \ee
and \be K_{\mu \nu} \stackrel{\rm def.}{=} \lambda \left[
\Phi_{\mu,\nu}-\Phi_{\nu,\mu} -\Phi_\rho \left({\gamma^\rho}_{\mu
\nu}-{\gamma^\rho}_{\nu \mu} \right)+ {{\gamma_{\mu
\nu}}^{\rho}}_{;\rho} \right], \ee and they are symmetric and skew
symmetric tensors, respectively. The energy-momentum tensor
$T^{\mu \nu}$ is given by  \be T^{\mu \nu}=g_{\rho \sigma}F^{\mu
\rho}F^{\nu \sigma}-\displaystyle{1 \over 4} g^{\mu \nu}
g^{\lambda \rho} g^{\epsilon \sigma} F_{\lambda \epsilon} F_{\rho
\sigma} \ee

\newsection{Cylindrically symmetric solution}

The tetrad space having cylindrical symmetry is given by

 \be
\left({e_i}^\mu \right)= \left( \matrix{ A & 0 & 0 & 0
\vspace{3mm} \cr 0 & i B \sin\theta \cos\phi & \displaystyle{i B1
\over r}\cos\theta \cos\phi
 & -\displaystyle{i B2 \sin\phi \over r \sin\theta} \vspace{3mm} \cr
0 & i B \sin\theta \sin\phi & \displaystyle{i B1 \over
r}\cos\theta \sin\phi
 & \displaystyle{i B2 \cos\phi \over r \sin\theta} \vspace{3mm} \cr
0 & i B \cos\theta & -\displaystyle{i B1 \over r}\sin\theta  & 0
\cr } \right)\ee where $i=\sqrt{-1}$ to preserve Lorentz signature
and  $A$, $B$, $B1$ and $B2$ are unknown functions in r and
$\theta$.

Applying (19) to the field equations (12)$\sim$(14) we note that
the two tensors $H_{\mu \nu}$ and  $K_{\mu \nu}$ are vanishing
identically regardless of the value of the functions $A$, $B$,
$B1$ and  $B2$. Thus M\o ller field equations reduce for the
tetrad (19) to Einstein's equations. Applying  the following
transformation \be R=\displaystyle{r \over B}, \ee to tetrad (19)
we obtain  \be \left({e_i}^\mu \right)= \left( \matrix{ A & 0 & 0
& 0 \vspace{3mm} \cr 0 & i (1-r B') \sin\theta \cos\phi  &
\displaystyle{i B1 \over r B}\cos\theta \cos\phi
 & -\displaystyle{i B2 \sin\phi \over r B \sin\theta} \vspace{3mm} \cr
0 & i (1-r B') \sin\theta \sin\phi & \displaystyle{i B1 \over r
B}\cos\theta \sin\phi
 & \displaystyle{i B2 \cos\phi \over r B \sin\theta} \vspace{3mm} \cr
0 &  i (1-r B') \cos\theta & -\displaystyle{i B1 \over r B
}\sin\theta  & 0 \cr } \right),\ee with $B'=\displaystyle{dB(r)
\over r}$. Then the field equations (12)$\sim$(14) take the form
\ba \A \A -\kappa T_{0 0} =  G_{0 0}, \qquad -\kappa T_{1 1} =
G_{1 1}, \qquad  -\kappa T_{1 2}= G_{1 1}  \nonu
\A \A -\kappa T_{2 1} = G_{2 1}, \qquad -\kappa T_{2 2}=  G_{2 2},
 \qquad -\kappa T_{3 3} = G_{3 3} , \ea where $G_{\mu \nu}$ is
defined by (15).

Now we are going to find a special solution to the non linear
partial differential equations (22), when  \be A =\displaystyle{1
\over \Lambda}, \qquad  B = \displaystyle{1 \over 2} ln(4
\sqrt{\Lambda}), \qquad B1 = \displaystyle{ B \over \Lambda},
\qquad B2 = B \Lambda,
 \ee
  $\Lambda$ is defined as \be \Lambda=1+ \displaystyle{1 \over
 4} {B_0}^2 r^2 \sin^2 \theta.\ee

 The form of the vector potential $A_\mu$,  the antisymmetric electromagnetic
 tensor field  $F_{\mu \nu}$ and the stress energy-momentum tensor ${T_\mu}^\nu$ are given by
  \ba \A \A A_t(r)=B_0 r \cos \theta, \qquad  A_t(\theta)=B_0 r \cos \theta, \qquad
  {\bf F}=B_0 \left(\cos \theta dt\wedge dr+r\sin\theta d\theta \wedge dt\right),\nonu
\A \A  {T_0}^0=-{T_3}^3=\displaystyle{{B_0}^2 \over 8\pi
\Lambda^4}, \qquad {T_1}^1=-{T_2}^2=\displaystyle{{B_0}^2(1-2
\sin^2 \theta) \over 8\pi \Lambda^4}, \nonu
\A \A  {T_1}^2=r^2 {T_2}^1=\displaystyle{-{B_0}^2 \sin \theta \cos
\theta \over 8\pi r \Lambda^4},
  \ea
  and the tetrad (21) takes the form
\be \left({e_i}^\mu \right)= \left( \matrix{ \Lambda & 0 & 0 & 0
\vspace{3mm} \cr 0 & \displaystyle{i \sin\theta \cos\phi \over
\Lambda} & \displaystyle{i \cos\theta \cos\phi \over r \Lambda}
 & -\displaystyle{i \Lambda \sin\phi  \over r  \sin\theta} \vspace{3mm} \cr
0 & \displaystyle{i \sin\theta \sin\phi \over \Lambda} &
\displaystyle{i \cos\theta \sin\phi \over r \Lambda}
 & \displaystyle{i \Lambda \cos\phi  \over r  \sin\theta} \vspace{3mm} \cr
0 &  \displaystyle{i \cos\theta \over \Lambda}   &
-\displaystyle{i \sin\theta \over r \Lambda } & 0 \cr }
\right),\ee where $\Lambda$ is given by (24) with the associated
Riemannian metric \be ds^2=\Lambda^2 \left(dt^2-dr^2 -r^2
d\theta^2 \right)-\displaystyle{r^2 \sin^2\theta  \over \Lambda^2}
d\phi^2, \ee  which is the line-element given before by Melvin
\cite{Mm}.

Thus we have an exact cylindrically symmetric solution satisfy the
field equations (12)$\sim$(14) and leads to the metric given by
Melvin in spherical  polar coordinate.

In what follows we will examine the singularities of solution (23)
and then, calculated the energy using the superpotential derived
form M\o ller's theory by Mikhail et al. \cite{MWHL}.

\newsection{Singularities}

In teleparallel theories we mean by singularity of spacetime
\cite{KT} the singularity of the scalar concomitants  of the
torsion and curvature tensors.

Using the definitions  of the  Riemann-Christoffel curvature
tensor, Ricci tensor, Ricci scalar, torsion tensor, basic vector,
traceless part and the axial vector part \cite{Nc2}, we obtain the
scalars of (23) in the form  \ba
 R^{\mu \nu \lambda \sigma}R_{\mu \nu \lambda
\sigma} \A = \A \displaystyle{ {B_0}^4  \over 4\Lambda^8} \left[
80-24{B_0}^2r^2 \sin^2 \theta+3 {B_0}^4r^4 \sin^4 \theta \right],
\nonu
R^{\mu \nu}R_{\mu \nu} \A=\A  \displaystyle{ 4{B_0}^4  \over
\Lambda^8} , \nonu
R \A=\A 0,\nonu
T^{\mu \nu \lambda}T_{\mu \nu \lambda} \A=\A  \displaystyle{
{B_0}^4r^2\sin^2 \theta  \over
2048\Lambda^4}\left[r^8{B_0}^8\sin^8
 \theta+24r^6{B_0}^6\sin^6  \theta +272r^4{B_0}^4\sin^4  \theta +
 1536r^2{B_0}^2\sin^2\theta+6144 \right], \nonu
\Phi^\mu \Phi_\mu \A=\A \displaystyle{ {B_0}^8r^6 \over
4096\Lambda^4}\left[-r^4{B_0}^4\sin^{10}
 \theta -24r^2{B_0}^2\sin^8  \theta -144\sin^6  \theta \right],\nonu
t^{\mu \nu \lambda}t_{\mu \nu \lambda} \A=\A \displaystyle{
{B_0}^4r^2 \over 4096\Lambda^4}\left[-r^8{B_0}^8\sin^{10}
 \theta-24r^6{B_0}^6\sin^8  \theta -336r^4{B_0}^4\sin^6  \theta -
2304r^2{B_0}^2\sin^4  \theta -9216\sin^2 \theta \right],\nonu
a^\mu a_\mu\A =\A 0.
 \ea
 As is clear from (28) that all the above scalars have a
 singularity when $\Lambda=0$. Also the scalars of Riemann-Christoffel curvature
tensor and Ricci tensor will approach infinity  rapidly than that
of torsion tensor, basic vector and traceless part when
$\Lambda\rightarrow 0$.
\newsection{The Energy content using M\o ller definition }

The superpotential of the M\o ller theory was given by Mikhail et
al. \cite{MWHL} as \be {{\cal U}_\mu}^{\nu \lambda} ={(-g)^{1/2}
\over 2 \kappa} {P_{\chi \rho \sigma}}^{\tau \nu \lambda}
\left[\Phi^\rho g^{\sigma \chi} g_{\mu \tau}
 -\lambda g_{\tau \mu} \gamma^{\chi \rho \sigma}
-(1-2 \lambda) g_{\tau \mu} \gamma^{\sigma \rho \chi}\right], \ee
where ${{\cal U}_\mu}^{\nu \lambda}$ is the superpotential derived
from M\o ller's theory and  ${P_{\chi \rho \sigma}}^{\tau \nu
\lambda}$ is  defined as \be {P_{\chi \rho \sigma}}^{\tau \nu
\lambda} \stackrel{\rm def.}{=} {{\delta}_\chi}^\tau {g_{\rho
\sigma}}^{\nu \lambda}+ {{\delta}_\rho}^\tau {g_{\sigma
\chi}}^{\nu \lambda}- {{\delta}_\sigma}^\tau {g_{\chi \rho}}^{\nu
\lambda} \ee with ${g_{\rho \sigma}}^{\nu \lambda}$ being a tensor
defined by \be {g_{\rho \sigma}}^{\nu \lambda} \stackrel{\rm
def.}{=} {\delta_\rho}^\nu {\delta_\sigma}^\lambda-
{\delta_\sigma}^\nu {\delta_\rho}^\lambda. \ee The energy is
expressed by the surface integral \cite{Mo2} \be E=\lim_{r
\rightarrow \infty}\int_{r=constant} {{\cal U}_0}^{0 \alpha}
n_\alpha dS, \ee where $n_\alpha$ is the unit 3-vector normal to
the surface element ${\it dS}$.

Now we are in a position to calculate the energy associated with
 solution (23) using  superpotential (29). As is
 clear from (32), the only components which contributes to the energy is ${{\cal U}_0}^{0
 \alpha}$. Thus substituting from (23) into
 (29) we obtain the following non-vanishing value
 \be
{{\cal U}_0}^{0 1}= \displaystyle{{B_0}^2
x\left[{B_0}^2(x^2+y^2)+8 \right] \over 128\pi}, \qquad {{\cal
U}_0}^{0 2}= \displaystyle{{B_0}^2 y\left[{B_0}^2(x^2+y^2)+8
\right] \over 128\pi}, \qquad  {{\cal U}_0}^{0 3}=0.
 \ee
 Substituting from (33) into
(32) we get \be E(r)=\displaystyle{1 \over 6}{B_0}^2
r^3+\displaystyle{1 \over 60}{B_0}^4 r^5. \ee This result depends
on the constant $B_0$ as to be expected.

\newsection{Main results and Discussion}

In this paper we have studied a cylindrically symmetric solution
in the teleparallel theory of gravitation and electromagnetic
fields. The axial vector part of the torsion, $a^\mu$ of this
solution is identically vanishing.

The solution gives rise to the same Riemannian metric given before
by Melvin \cite{Mm, Mm1}.  Melvin  presented a rigorous solution
of Einstein-Maxwell equations which correspond to a configuration
of parallel magnetic lines of force in equilibrium under their
mutual gravitational attraction.  This solution had been obtained
earlier by Misra and Radhakrishna \cite{MR}.  This spacetime is
invariant under rotation about and translation along an axis of
symmetry. This is also invariant under reflection in planes
comprising that axis or perpendicular to it. Wheeler \cite{Wj}
demonstrated  that a magnetic universe could also be obtained in
Newton's theory of gravitation and showed that it is unstable
according to the elementary Newtonian analysis. Further Melvin
\cite{Mm1}  showed that his universe to be  stable against small
radial perturbation and Thorne \cite{Tk} proved the stability of
the magnetic universe against arbitrary large perturbation.
Further Thorne \cite{Tk} pointed out that the Melvin magnetic
universe might be of great value in understanding the nature of
extragalactic sources of radio waves and thus the Melvin solution
to the Einstein-Maxwell equations is of immense astrophysical
interest. Virbhadra and Prasanna \cite{VP} studied spin dynamics
of charged massive test particles in this spacetime. Energy
distribution in Melvin's universe computed by many authors
\cite{Xi,RC} using different definitions of the energy momentum
complex within the framework of general relativity theory.

It was shown by M\o ller \cite{Mo3} that a tetrad description of a
gravitational field equation allows a more satisfactory treatment
of the energy-momentum complex than does general relativity.
Therefore, we have applied the superpotential method given by
Mikhail et al.\cite{MWHL}   to calculate the energy.  As is clear
from equation (34) that energy  depends on the constant magnetic
$B_0$ and if this constant equal zero then, $E=0$ since solution
(23) will reduce to a flat space time i.e.
\[ds^2=dt^2-dr^2-r^2(d\theta^2-\sin \theta^2 d\phi^2). \]

 It is of interest  to mention that Radinschi \cite{RC3} has
calculated the energy distribution of Melvin metric using M\o ller
energy momentum complex within the framework of general relativity
and obtained  \be E(r)=\displaystyle{1 \over 3} {B_0}^2
r^3-\displaystyle{1 \over 15} {B_0}^4 r^5+\displaystyle{1 \over
70} {B_0}^6 r^7.\ee Also Xulu \cite{Xi} has calculated the energy
of the same metric using the energy momentum complex given by
Landau and Lifshitz and the energy momentum complex given by
Papapetrou and obtained  \be E(r)=\displaystyle{1 \over 6} {B_0}^2
r^3+\displaystyle{1 \over 20} {B_0}^4 r^5+\displaystyle{1 \over
140} {B_0}^6 r^7+\displaystyle{1 \over 2520} {B_0}^8 r^9.\ee As is
clear that there is a difference between the results obtained by
Radinschi,  Xulu and the results obtained here. This is expected
because as we stated above that general relativity theory has a
problem in the calculation of the quasi local energy
\cite{Vs,Vs1}. This problem is not exists in the teleparallel
theories because it has only one definition of the energy-momentum
complex \cite{Mo1}.
\newpage
A comparison between the results of the energy using different
energy momentum complexes are given in the following table
\begin{center}
Table (I)  comparison between the results of energy using
different energy momentum complexes
\end{center}
\begin{center}
\begin{tabular}{|c|c|c|} \hline
 & & \\ Energy Momentum Complex  & Energy&Equation \\ & &
\\ \hline M\o ller & & \\ Within the
Framework of & $\displaystyle{1 \over 6}{B_0}^2
r^3+\displaystyle{1 \over 60}{B_0}^4 r^5$ & (34) \\ Tetrad
Spacetime & & \\ \hline  M\o ller & &
\\ Within the Framework of & $\displaystyle{1 \over 3} {B_0}^2
r^3-\displaystyle{1 \over 15}
{B_0}^4 r^5+\displaystyle{1 \over 70} {B_0}^6 r^7$&(35)  \\
 General Relativity Spacetime & &\\
\hline Landau
and Lifshitz, Papapetrou& & \\
 Within the
Framework of &$\displaystyle{1 \over 6} {B_0}^2
r^3+\displaystyle{1 \over 20} {B_0}^4 r^5+\displaystyle{1 \over
140} {B_0}^6 r^7+\displaystyle{1 \over 2520} {B_0}^8 r^9$&(36) \\
General Relativity Spacetime
& &\\
\hline
\end{tabular}
\end{center}

\end{document}